\documentclass[10pt,twocolumn,
superscriptaddress,showkeys,
 amsmath,amssymb,
 aps,
pra,
]{revtex4-2}
\usepackage{graphicx}
\usepackage{dcolumn}
\usepackage{bm}
\usepackage{xcolor}
\usepackage{soul}
\usepackage{multirow}
\usepackage[mathlines]{lineno}
\usepackage[colorlinks,linkcolor=blue,anchorcolor=blue,citecolor=blue,]{hyperref}
\newcommand*{\wn}{cm$^{-1}$}
\usepackage[greek,english]{babel}
\newcommand{\XS}{$X^2\Sigma^+$}
\newcommand{\Ab}{$A^2\Pi_{3/2}$}
\newcommand{\Aa}{$A^2\Pi_{1/2}$}
\newcommand{\AS}{$A^2\Pi$}

\begin{document}

\title{Precision spectroscopy of the \AS\  $\leftarrow$ \XS\ transition in BaF} 

\author{M.C. Mooij}
 \affiliation{Department of Physics and Astronomy, LaserLaB, Vrije Universiteit \\
 De Boelelaan 1100, 1081 HZ Amsterdam, The Netherlands}
 \affiliation{Nikhef, National Institute for Subatomic Physics, Science Park 105, 1098 XG Amsterdam, The Netherlands}

\author{H.L. Bethlem}
 \affiliation{Department of Physics and Astronomy, LaserLaB, Vrije Universiteit \\
 De Boelelaan 1100, 1081 HZ Amsterdam, The Netherlands}
 \affiliation{Nikhef, National Institute for Subatomic Physics, Science Park 105, 1098 XG Amsterdam, The Netherlands}

\author{W. Ubachs}\thanks{Corresponding author; email: w.m.g.ubachs@vu.nl}
 \affiliation{Department of Physics and Astronomy, LaserLaB, Vrije Universiteit \\
 De Boelelaan 1100, 1081 HZ Amsterdam, The Netherlands}

\author{P. Aggarwal} 
 \affiliation{Van Swinderen Institute for Particle Physics and Gravity (VSI), University of Groningen, Nijenborgh 3, 9747 AG Groningen, The Netherlands}
 \affiliation{Nikhef, National Institute for Subatomic Physics, Science Park 105, 1098 XG Amsterdam, The Netherlands} 

\author{A. Boeschoten} 
 \affiliation{Van Swinderen Institute for Particle Physics and Gravity (VSI), University of Groningen, Nijenborgh 3, 9747 AG Groningen, The Netherlands}
 \affiliation{Nikhef, National Institute for Subatomic Physics, Science Park 105, 1098 XG Amsterdam, The Netherlands}  

\author{A. Borschevsky}
 \affiliation{Van Swinderen Institute for Particle Physics and Gravity (VSI), University of Groningen, Nijenborgh 3, 9747 AG Groningen, The Netherlands}
 \affiliation{Nikhef, National Institute for Subatomic Physics, Science Park 105, 1098 XG Amsterdam, The Netherlands}  

\author{Y. Chamorro} 
\affiliation{Van Swinderen Institute for Particle Physics and Gravity (VSI), University of Groningen, Nijenborgh 3, 9747 AG Groningen, The Netherlands}

\author{M. Denis} 
\affiliation{Van Swinderen Institute for Particle Physics and Gravity (VSI), University of Groningen, Nijenborgh 3, 9747 AG Groningen, The Netherlands}
 \affiliation{Nikhef, National Institute for Subatomic Physics, Science Park 105, 1098 XG Amsterdam, The Netherlands}

\author{T.H. Fikkers} 
 \affiliation{Van Swinderen Institute for Particle Physics and Gravity (VSI), University of Groningen, Nijenborgh 3, 9747 AG Groningen, The Netherlands}
 \affiliation{Nikhef, National Institute for Subatomic Physics, Science Park 105, 1098 XG Amsterdam, The Netherlands}

\author{S. Hoekstra}
 \affiliation{Van Swinderen Institute for Particle Physics and Gravity (VSI), University of Groningen, Nijenborgh 3, 9747 AG Groningen, The Netherlands}
 \affiliation{Nikhef, National Institute for Subatomic Physics, Science Park 105, 1098 XG Amsterdam, The Netherlands}

\author{J.W.F. van Hofslot}
 \affiliation{Van Swinderen Institute for Particle Physics and Gravity (VSI), University of Groningen, Nijenborgh 3, 9747 AG Groningen, The Netherlands}
 \affiliation{Nikhef, National Institute for Subatomic Physics, Science Park 105, 1098 XG Amsterdam, The Netherlands}

\author{S.A. Jones} 
\affiliation{Van Swinderen Institute for Particle Physics and Gravity (VSI), University of Groningen, Nijenborgh 3, 9747 AG Groningen, The Netherlands}
 \affiliation{Nikhef, National Institute for Subatomic Physics, Science Park 105, 1098 XG Amsterdam, The Netherlands}

\author{V.R. Marshall} 
 \affiliation{Van Swinderen Institute for Particle Physics and Gravity (VSI), University of Groningen, Nijenborgh 3, 9747 AG Groningen, The Netherlands}
 \affiliation{Nikhef, National Institute for Subatomic Physics, Science Park 105, 1098 XG Amsterdam, The Netherlands}

\author{T.B. Meijknecht}
\affiliation{Van Swinderen Institute for Particle Physics and Gravity (VSI), University of Groningen, Nijenborgh 3, 9747 AG Groningen, The Netherlands}
 \affiliation{Nikhef, National Institute for Subatomic Physics, Science Park 105, 1098 XG Amsterdam, The Netherlands}

\author{R.G.E. Timmermans}
\affiliation{Van Swinderen Institute for Particle Physics and Gravity (VSI), University of Groningen, Nijenborgh 3, 9747 AG Groningen, The Netherlands}
 \affiliation{Nikhef, National Institute for Subatomic Physics, Science Park 105, 1098 XG Amsterdam, The Netherlands}

\author{J. de Vries}
\affiliation{Institute for Theoretical Physics, University of Amsterdam, Science Park 904, 1098 XH Amsterdam, The Netherlands}
 \affiliation{Nikhef, National Institute for Subatomic Physics, Science Park 105, 1098 XG Amsterdam, The Netherlands}
 
\author{L. Willmann}
\affiliation{Van Swinderen Institute for Particle Physics and Gravity (VSI), University of Groningen, Nijenborgh 3, 9747 AG Groningen, The Netherlands}
 \affiliation{Nikhef, National Institute for Subatomic Physics, Science Park 105, 1098 XG Amsterdam, The Netherlands}

\author{the NL-$e$EDM collaboration}
 
\date{\today}

\begin{abstract}
\noindent
High-resolution spectroscopy on the \AS\ - \XS\ electronic system of $^{138}$Ba$^{19}$F is performed using a cold molecular beam produced by a buffer gas source.
The hyperfine structure in both \XS\ ground and \AS\ excited states is fully resolved and absolute transition frequencies of individual components are measured at the sub-MHz level making use of frequency-comb laser calibration. 
Sets of molecular constants for the \XS($v=0,1$) and \AS($v=0,1$) levels are determined, with improved accuracy for the $T_{v',v''}$ band origins and spin-orbit interaction constants for the \AS\ excited states, that represent the presently measured highly accurate transitions for low-$J$ states as well as previously determined transition frequencies in Fourier-transform emission studies for rotational levels as high as $J \geq 100$.
The extracted molecular constants reproduce the measured transition frequencies at the experimental absolute accuracy of 1 MHz.
The work is  of relevance for future laser cooling schemes, and is performed in the context of a measurement of the electron dipole moment for which BaF is a target system.
\end{abstract}

\maketitle

\section{Introduction}
\label{intro}

The spectroscopy of the barium monofluoride molecule in the gas phase has attracted renewed interest since it was considered as a target molecule for detecting an electric dipole moment of the electron ($e$EDM)~\cite{Aggarwal2018}. 
The molecular enhancement factors for the competing molecular systems from which tight constraints on an $e$EDM have already been extracted  (YbF~\cite{Hudson2011}, ThO~\cite{Andreev2018}, and HfF$^+$~\cite{Roussy2023}) are larger than that of BaF~\cite{Haase2021}. 
However, the prospects that the electronic structure of BaF~\cite{Kang2016,Hao2019} provide for Stark deceleration~\cite{Aggarwal2021b}, hexapole focusing~\cite{Touwen2024}, laser cooling~\cite{Chen2017,Rockenhauser2024,Kogel2025,vanHofslot2025},  trapping~\cite{Zeng2024}, and entraining in a solid matrix~\cite{Vutha2018a,Messineo2024} make this molecule a good candidate for an $e$EDM search.
Spin precession methods for detecting an $e$EDM, for which only the $F=0$ and $F=1$ hyperfine components and their $M_F$ Zeeman sub-components within the $N=0, J=1/2$ ground rotational level of BaF are used, have already been demonstrated~\cite{Boeschoten2024}.

The overall electronic structure of the lowest states in the BaF molecule and their molecular constants were determined in the 1990s by a Lyon-Orsay-Oxford collaboration in a sequence of spectroscopic works~\cite{Effantin1990,Bernard1990,Bernard1992}.
Subsequently, the vibrational structure in the \XS\ electronic ground state of BaF was investigated via Fourier-transform emission spectroscopy~\cite{Guo1995}.
The hyperfine structure in the \XS\ ground vibrational level was investigated via rotational spectroscopy~\cite{Ryzlewicz1980} and microwave-optical double resonance spectroscopy~\cite{Ernst1986}, in combination with delivering accurate constants for the magnetic dipole hyperfine interactions of the F-nucleus in the $^{138}$Ba$^{19}$F molecule.

The \AS-\XS(0,0) band was studied by Steimle et al.~\cite{Steimle2011} in a high-resolution laser-induced fluorescence spectroscopy experiment. 
This work delivered, besides field-free measurements, a set of dipole moments for the \Aa\ and \Ab\ spin-orbit states and magnetic $g$-factors for the \XS$(v=0)$ and \AS$(v=0)$ states.
From the spectroscopy an improved set of molecular constants were produced, while achieving a resolution of 30 MHz and a spectroscopic accuracy of 21 MHz as the standard deviation in their fit to the data. 
While spin-rotation splittings were resolved, the resolution was insufficient to resolve hyperfine structure.
The hyperfine structure of a transition useful for laser cooling was resolved in saturated absorption from a buffer gas source~\cite{Bu2022,Zhang2022}, but no absolute frequencies were reported.
Similarly, Langen and coworkers developed a buffer gas source of BaF and performed absorption and fluorescence imaging spectroscopic studies of $^{138}$BaF and $^{136}$BaF with the goal to characterize the molecular structure for applying laser cooling to this molecule~\cite{Albrecht2020,Rockenhauser2023}.

The present study reports on a renewed high-resolution investigation of the \AS-\XS\ system of BaF.
Within the framework of the NL-$e$EDM collaboration, an experimental setup was built based on a buffer-gas molecular beam source for BaF molecules~\cite{Mooij2024,Mooij2025}.
Hyperfine-resolved transition frequencies were determined for a number of lines in the \AS-\XS\ system, probing both \Aa\ and \Ab\ spin-orbit states and (0,0), (1,1) and (2,2) bands.
A fit to the absolutely calibrated transition frequencies is performed to extract molecular constants for the rotational structure in the \AS\ state for ($v=0$) and ($v=1$).

\section{Experimental setups}
\label{exp}

\begin{figure}[b]
    \centering
    \includegraphics[width=0.98\columnwidth]{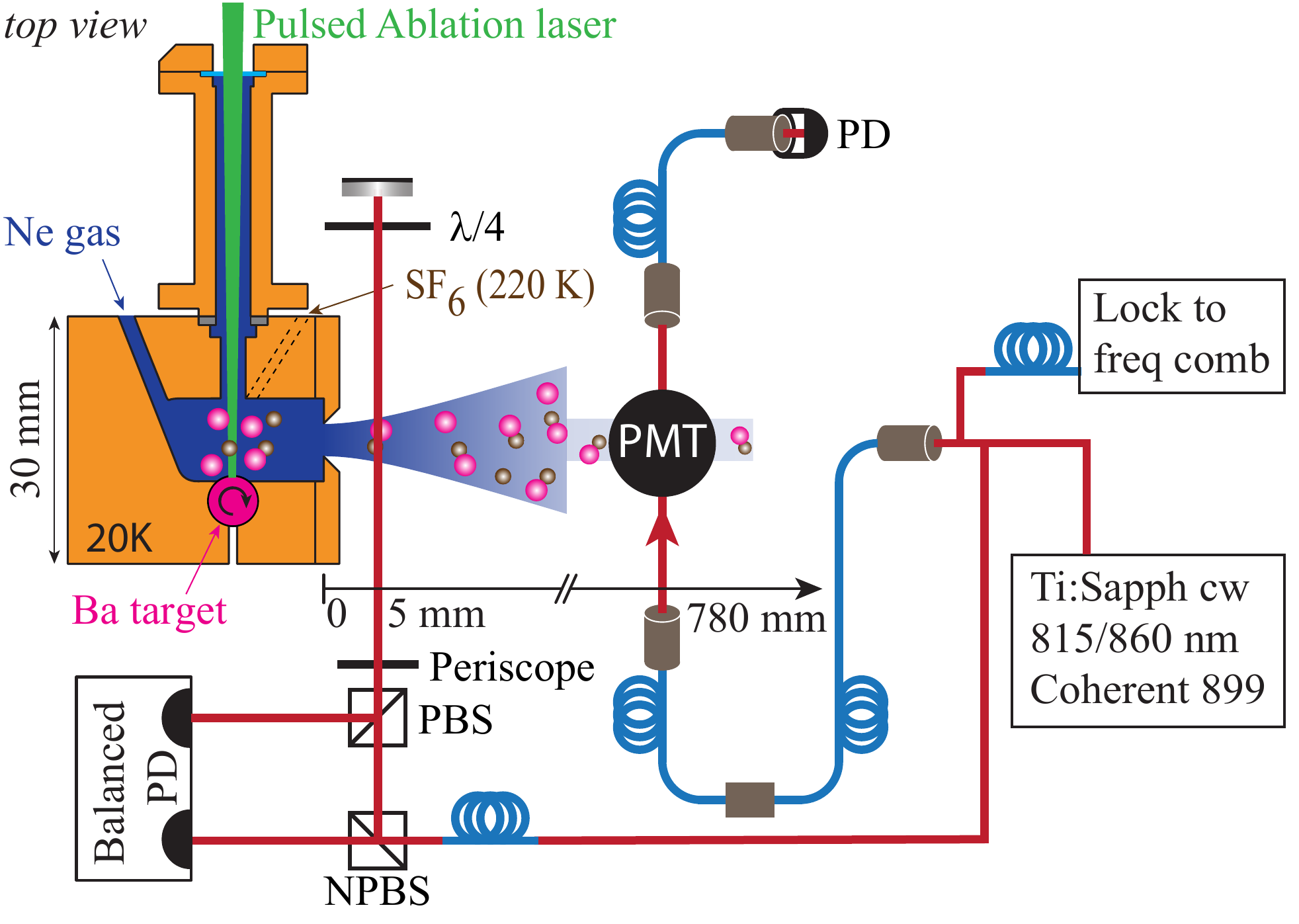}
    \caption{Layout of the the experimental setup showing the cryogenic buffer-gas cooled beam source for BaF and the two detection zones for recording spectra. At a distance of 5 mm from the source exit, spectra are recorded in absorption from a two-way pass through the molecular beam. At a distance of 780\;mm, laser-induced fluoresence is detected for investigating the \AS\ $\leftarrow$ \XS\ transitions, exciting with both 860 nm and 815 nm laser radiation. In this region Doppler shifts are assessed and minimized using overlapping counter-propagating laser beams. For further details see main text and Refs.~\cite{Mooij2024,Mooij2025,MooijThesis}.
   }
    \label{fig:setup}
\end{figure}

A buffer gas source to generate a cold beam of BaF molecules was built at LaserLaB Amsterdam for producing a beam with a typical velocity of 200 m/s.
For its construction a design by Truppe et al.~\cite{Truppe2018} and a prototype source built at University of Groningen was followed~\cite{Esajas2021-PhD}. 
The source is presented in schematic form in Fig.~\ref{fig:setup}.
The buffer gas source is operated at a repetition rate of 10 Hz. 
For detailed studies characterizing the source and the molecular beam produced, including its longitudinal phase-space distribution and the measurement of its brightness at $1.3 (5) \times 10^{11}$ molecules per steradian per pulse in $N=0$, we refer to previous studies~\cite{Mooij2024,Mooij2025}.
In the setup two spectroscopic zones are defined.

At 5 mm behind the source exit a laser beam is crossed perpendicularly to the molecular beam to perform a direct absorption experiment. The laser beam is retro-reflected to double the signal, making use of balanced photodiodes and dividing out the laser intensity. 
In this geometry a typical Doppler width of 120 MHz, corresponding to an average transversal velocity of 100 m/s, is observed.

In a second interaction zone, at a distance of 780 mm from the source exit, laser-induced fluorescence (LIF) is collected. 
Titanium-sapphire (Ti:Sa) lasers were operated in continuous-wave mode for detection, with signal averaging over the duration of the molecular pulses in the interaction zone.
Skimming and collimation of the molecular beam, combined with detection by perpendicularly crossing laser beams in the far field, yields a small Doppler broadening of some 2 MHz, even smaller than the contribution of natural lifetime broadening, which is 2.8 MHz for the \Aa\ state and 3.3 MHz for the \Ab\ state as derived from lifetime measurements~\cite{Aggarwal2019}. 
A typical resolution at or below 5 MHz is achieved, sufficient to resolve the hyperfine structure in $A-X$ transitions, at least for the low rotational angular momentum states. 
Retro-reflection of the laser-beams is used, in an off-line experiment, to assess the small Doppler shift resulting from a non-exact perpendicular alignment of the lasers with respect to the molecular beam.
This procedure leads to a reduction of the first order Doppler shift to 10 kHz and to a high absolute accuracy in the frequency determination of the hyperfine resolved resonances.

The photon collection efficiency in the laser-induced fluorescence zone in the far field is about 0.06.
The quantum efficiency of the PMT probing the fluorescence is $< 0.1\%$ at 860 nm and 1\% at 815 nm, hence excitation of the higher-lying \Ab\ states yields a better signal-to-noise ratio than the \Aa\ states.
The vibrational temperature in the beam depends on the buffer gas flow rates, yielding $T_{\rm vib}=408$ K for 20 \textsc{sccm} and $T_{\rm vib}=304$ K for 70 \textsc{sccm}. This makes it possible to probe $v=1$ and $v=2$ vibrational states. 
Rotational temperatures vary and are also dependent on the flow rate in the buffer gas cell, but are typically below 10 K.

A significant improvement to previous spectroscopic studies on BaF is the locking of the Ti:Sa lasers via in-house fiber-connections to a frequency-comb laser (Menlo Systems, FC1500-250-WG), which itself is locked to the in-house Cs-clock (Microsemi CSIII). Via beat-note measurements of the continuous-wave Ti:Sa lasers against the frequency comb a sub-MHz absolute accuracy of the frequency scale is obtained.

\section{Overview spectra} 

\begin{figure}[htb]
    \centering
    \includegraphics[width=\linewidth]{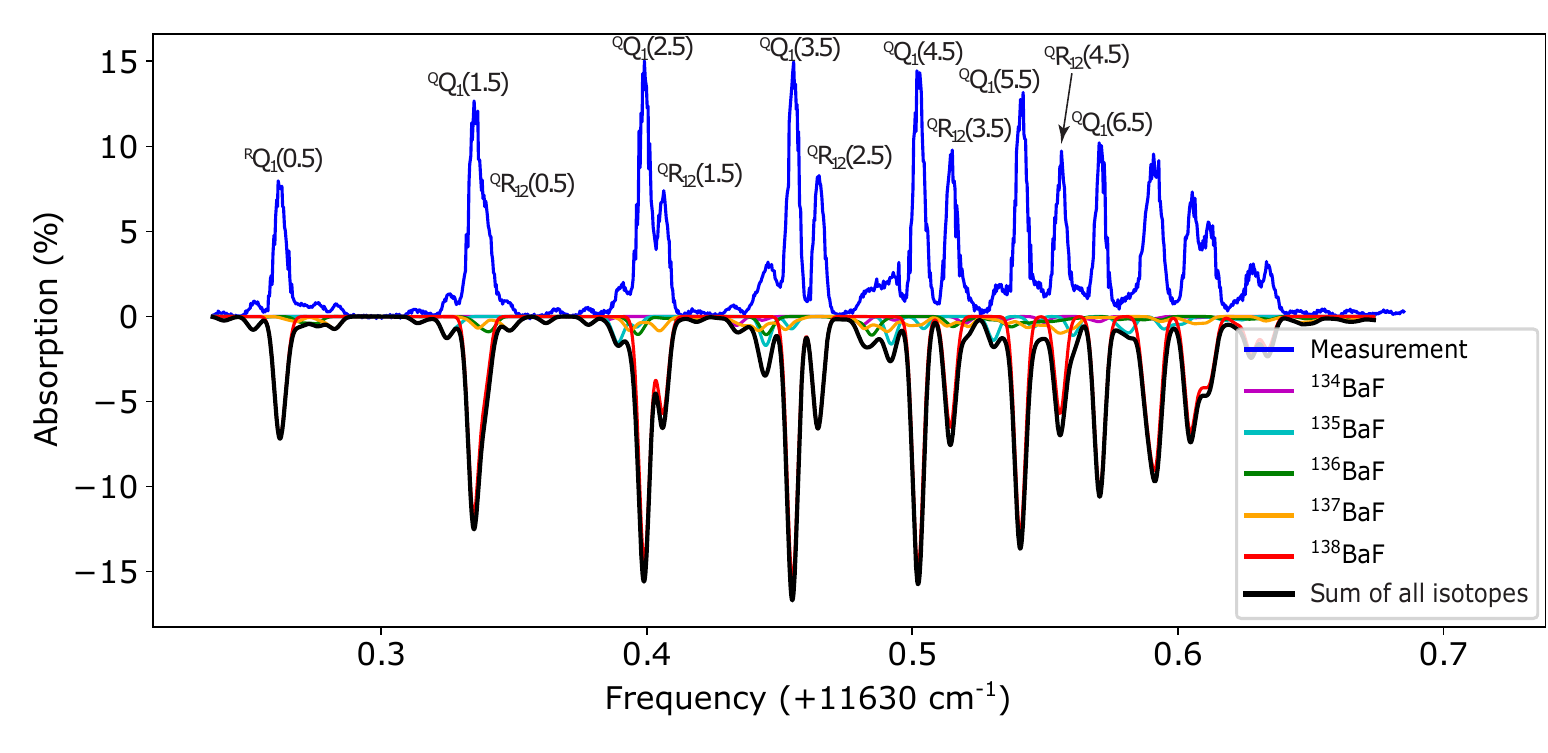}
    \caption{Absorption measurement (above) and simulation (below) of the $^QQ$ and $^QR$ rotational branches of the \Aa\ $\leftarrow$ \XS(0,0) band of BaF. The absorption intensity is the time averaged signal over the duration of the pulse.  The simulation includes the five strongest isotopes shown in different colors as indicated in the legend. The black curve shows the combined spectrum.
    }
    \label{fig:absSpectrum}
\end{figure}

The absorption zone at 5 mm behind the exit of the cryogenic buffer gas source was used for the measurement of an overview spectrum of two branches of the \Aa\ $\leftarrow$ \XS(0,0) band which is presented in Fig.~\ref{fig:absSpectrum}. 
The linewidths obtained in these experiments in the near-field are Doppler-limited at $\sim 120$ MHz.
The measured $^QQ$- and $^QR$-branches are depicted, which are nested into each other. 
A simulated spectrum is also shown, including the contribution of the five most dominant isotopes, taking into account their natural abundances. 

The simulation is performed using \textsc{pgopher}~\cite{Western2017}, a program for simulating rotational, vibrational and electronic spectra~\cite{Western2017}.
A rotational temperature of 12\;K is extracted. 
The molecular constants for $^{138}$BaF, $^{137}$BaF, and $^{135}$BaF are taken from Ref.~\cite{Steimle2011}. 
For $^{136}$BaF and $^{134}$BaF, the molecular constants were not available at sufficient accuracy and therefore estimated via isotopic scaling. 
In the present spectroscopic study, the focus is entirely on $^{138}$BaF.

Although generally the simulated spectrum agrees very well with the measurements, it may be seen in Fig.~\ref{fig:absSpectrum} that the experimental values for the intensity ratios between the $^QR$ and $^QQ$ branches deviate from those predicted in simulation. 
Quantitative information and further details were given in a PhD Thesis~\cite{MooijThesis}. 
Such intensity anomalies and deviations from H\"onl-London factors may be explained by quantum interferences between electronic states of different symmetry. 
Recently, in an example of the observed anomalous intensities in the $C^2\Sigma^+$ - $X^2\Sigma^+$(0,0) band in RaOH was quantitatively attributed to mixing between excited $C^2\Sigma^+$ and \Aa\ states~\cite{Conn2025}. 
Also, in the example of $\ell$-uncoupling in Rydberg states of molecular nitrogen, where interference between $\Sigma$ and $\Pi$ symmetry states plays a role, intensity ratios between P and R-branches were found to strongly deviate from predicted H\"onl-London factors~\cite{Liu2006}. 
In a similar fashion the intensity anomalies detected in the \Aa\ - \XS(0,0) band of BaF might be attributed to interaction of the \Aa\ state with the $B^2\Sigma^+$ state, which lies at an excitation energy of $T_0 = 14040.163$ \wn~\cite{Effantin1990}.

\section{Precision spectroscopic results}

\begin{figure}
    \centering
    \includegraphics[width=0.96\linewidth]{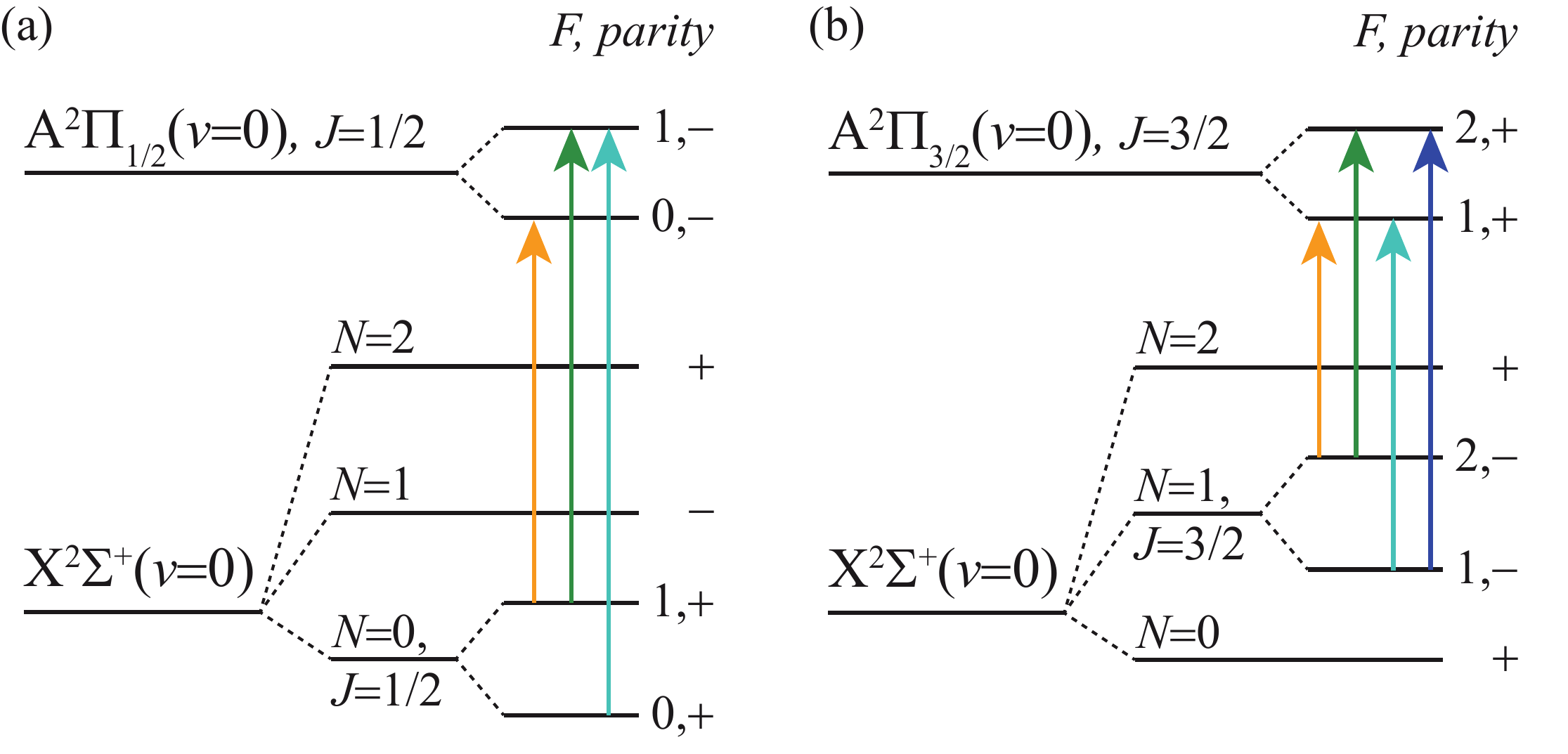}
    \caption{Level and transition schemes for the spectroscopic investigation of hyperfine-resolved LIF spectra of $^{138}$Ba$^{19}$F. Excitation from the $N=0, J=1/2$ ground state is depicted for:
    transition (a) to \Aa, $J=1/2$; (b) to \Ab, $J=3 /2$.
    In the experiments, excitation from higher lying rotational levels $N=1-5$ and from \XS, $v=1,2$ are additionally performed. The $\Lambda$-doubling in the \AS\ state is not shown.}
    \label{fig:levels}
\end{figure}

The high-resolution studies are performed by detecting laser-induced fluorescence in the far field of the collimated molecular beam.
The excitation schemes and the level structure of BaF with its characteristic spin-doublet structure is depicted in Fig.~\ref{fig:levels}. 
Characteristic for the spectrum is the splitting of each rotational level in the ground state by the spin-rotation interaction $\vec{N}\cdot\vec{S}$ and the subsequent level doubling by the magnetic dipole hyperfine interaction $\vec{J}\cdot\vec{I}$.
In the excited state the spin-orbit interaction splits the levels into \Aa\ and \Ab\ manifolds to be excited at largely differing wavelengths of 860 nm and 815 nm.
In the excited states of $\Pi$-symmetry each spin-orbit ladder is further split by the $\Lambda$-doubling producing two levels of opposite parity. 
These splittings range from 7723 MHz in the \Aa\ $J=1/2$ level and 0.015 MHz in the \Ab\ $J=3/2$ level to 1.3 MHz in the highest \Ab\ $J=15/2$ level excited.
Hence the $\Lambda$-doublet splittings in \Aa\ are several orders of magnitude larger than in the \Ab\ manifold.
Even if some of these splittings are smaller than the narrowest observed linewidths, they do not influence the spectroscopic analysis, since the components of a $\Lambda$-doublet exhibit opposite parity, and selection rules dictate that only one of the components is observed in each rotational transition.
Each resulting level in the $\Lambda$-doubled rotational levels is further split into two components by the $I=1/2$ nuclear spin of the $^{19}$F nucleus.

The focus is on the determination of absolute frequencies of the hyperfine-resolved resonances. 
All measurements are performed with online calibration via beat-note measurements against the frequency comb laser.
For the precision measurements laser-induced fluorescence (LIF) is collected for signal acquisition at large distance from the buffer gas exit (780 mm).
In this zone retro-reflection of the laser-beams, with in-coupling of the reflected beams into single-mode fibers, is used to verify the small Doppler shift resulting from a non-exact perpendicular alignment of the lasers with respect to the molecular beam.
From this procedure it is established that the angle between the two counter-propagating laser beams is below $5 \times 10^{-3}$ degrees, corresponding to a Doppler shift of below 10 kHz for the BaF molecular beam at 200 m/s~\cite{Mooij2025}.

\begin{figure*}[htb]
	\centering
	\includegraphics[width=0.8\linewidth]{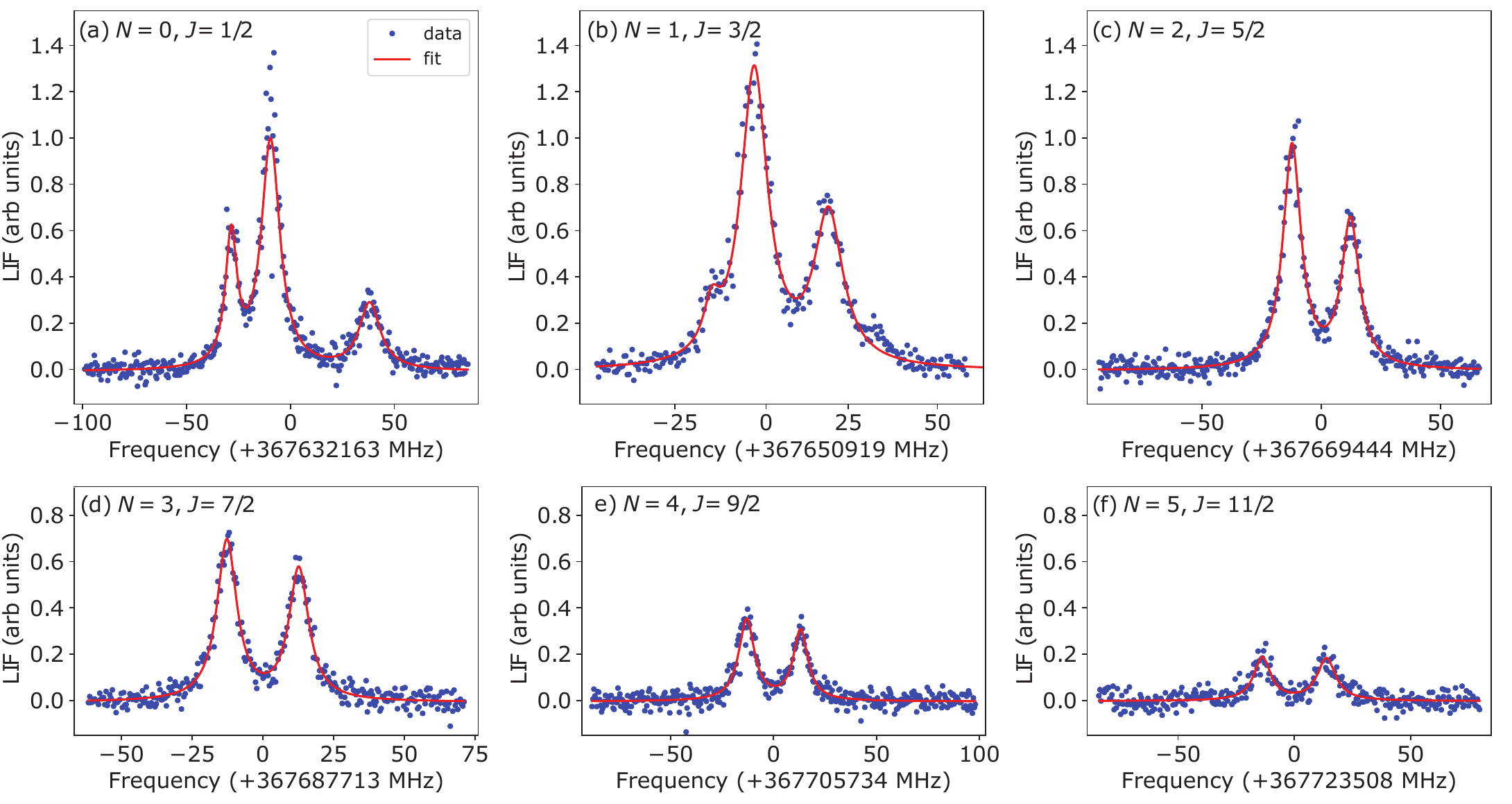}
	\caption{Spectra of the $^SR$-branch of the \Ab\ $\leftarrow$ \XS(0,0) band at $\lambda = 815$ nm with from (a) to (f) the transitions originating from the $N=0, J=1/2$ to $N=5, J=11/2$  rotational ground states, respectively. Spectra are measured at $z=780$\;mm from the buffer gas source exit using LIF detection. The frequency axis is online calibrated with respect to the frequency comb laser. Each data point originates from integration over a single molecular beam pulse while subtracting the integrated background signal observed during the same time interval after the molecular beam pulse has passed. Note that the intensity scale for the different panels (a) - (f) is the same, and the intensity decrease toward larger $N$ reflects the population of rotational states in the beam.
    The solid red curves show the result of the multi-component fit of Lorentzians to the observed data.}
	\label{fig:A32-X}
\end{figure*}

\begin{figure*}[htb]
	\centering
	\includegraphics[width=0.6\linewidth]{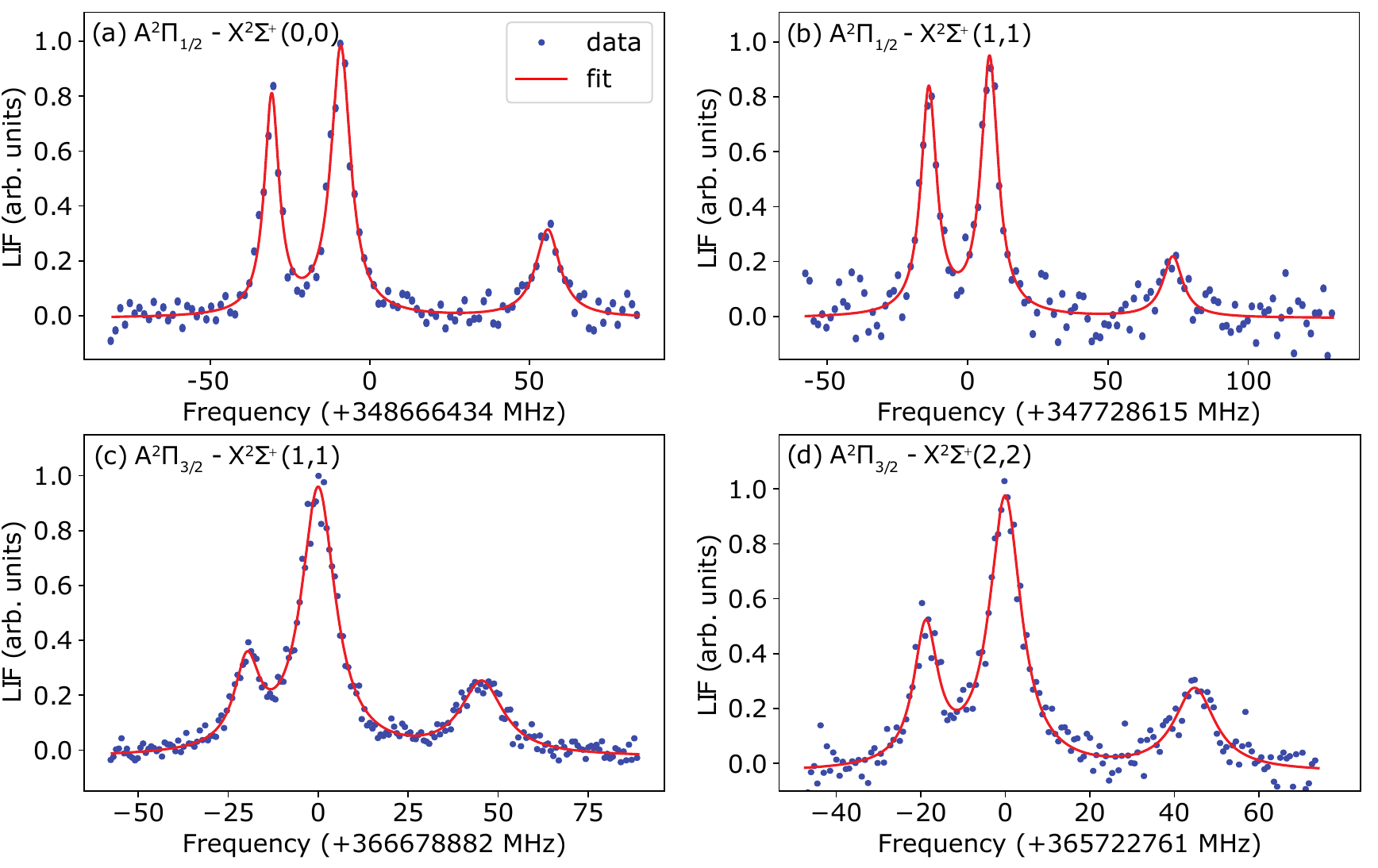}
	\caption{Recorded spectra of a number of transitions in various bands of BaF. 
    Transitions are in the $A-X$($v',v''$) bands, probe \Aa\ or \Ab\ spin-orbit components, and originate from the lowest $N=0,J=1/2$ levels in the $v''$ vibrational states.   
    The identification and extracted frequencies of hyperfine-resolved components for these lines are given in Table~\ref{tab42}. The spectrum shown in panel (a) results from averaging over 5 shots, while the spectra shown in panel (b) to (d) have been averaged over 20-25 shots. The recorded LIF intensity of the spectrum shown in panel (a) is about an order of magnitude smaller than that shown in Fig.~\ref{fig:A32-X}(a) due to the lower quantum efficiency of the PMT at 860\,nm compared to 815\,nm. The intensity of the spectrum shown in panel (b) is about two orders of magnitude smaller than that of the spectrum shown in (a) due to the lower population of the $v=1$ state. Similarly, the spectrum shown in panel (c) is about two orders of magnitude smaller than that of Fig.~\ref{fig:A32-X}(a), while the spectrum shown in panel (d) is about a factor of 5 less intense than the one shown in (c) due to the even lower population of the $v=2$ state.} 
	\label{fig:A12-X}
\end{figure*}

Typical spectra as recorded for a sequence of rotational lines in the $^SR$ branch of the \Ab\ $\leftarrow$ \XS(0,0) band of BaF are presented in Fig.~\ref{fig:A32-X}.
In these measurements the hyperfine structure was resolved, and the transition frequencies were determined at an absolute accuracy ranging from 0.2 MHz to 0.7 MHz depending on statistics and signal-to-noise ratio. 
Four measurements were performed, two from each side and two for different polarizations.
Measurements at perpendicular polarization geometries were carried out to assess possible effects of Zeeman shifts. 
Transition frequencies were also obtained from spectra taken with flow rates between 30 and 70\;\textsc{sccm} of the Ne buffer gas in the beam source. This led to no significant deviations. 

Similarly, a set of lines probing low-$N,J$ states in the $A-X$ system, measured in the (0,0) and (1,1) bands, as well as a line in the (2,2) band, are presented in Fig.~\ref{fig:A12-X}.
The potential energy curves in the electronic systems of BaF are known to favor diagonal vibrational transitions, and this holds also for the \AS\ $\leftarrow$ \XS\ system~\cite{Hao2019}, allowing for the observation of transitions in the (1,1) and (2,2) bands.
At vibrational temperatures equal or exceeding room temperature in the molecular beam the population of the $v=1$ state is effectively an order of magnitude weaker than that of $v=0$, while $v=2$ is found another factor of 5 weaker.

The transition frequencies resulting from the measurements, displayed in Figs.~\ref{fig:A32-X} and \ref{fig:A12-X}, are listed in Tables~\ref{tab41} and \ref{tab42}, respectively.
The contribution of Doppler broadening in the LIF zone is reduced to a minimum and the spectral components are well described by Lorentzian profiles.
The resonance frequencies are found by fitting multi-Lorentzian functions to the fluorescence data. 
As is apparent from Figs.~\ref{fig:A32-X} and \ref{fig:A12-X} for the transitions combining the lowest $N,J$ rotational levels, all three possible hyperfine sub-components remain resolved, and their three separate transition frequencies are listed in Tables~\ref{tab41} and \ref{tab42}.
For transitions connecting higher $J$ rotational levels the splittings between $F=J \pm 1/2$ hyperfine subcomponents in the excited \Ab\ state tend to become smaller, so that the threefold-resolved structure  transfers into a two-component structure.
The sequence of panels (a) - (c) in Fig.~\ref{fig:A32-X}
illustrates this. The shoulder in panel \ref{fig:A32-X}(b) reveals a third hyperfine component that is clearly separated in panel \ref{fig:A32-X}(a), and is no longer visible in panel \ref{fig:A32-X}(c).
For the $\Delta J=1$ transitions in the $^S R$-branch the $\Delta F=0$ components tend to become much weaker than the $\Delta F=1$ components, due to $6J$-symbols governing the intensities.
For this reason the assignments as listed in Table~\ref{tab41} relate to the resolved strong $\Delta F=1$ components.

\begin{table*}
    \renewcommand{\tabcolsep}{1pt}
    \centering
    \caption{Measured transition frequencies between the $X^2\Sigma^+,v=0$ and the $A^2\Pi_{3/2},v=0$ manifolds starting from a sequence of rotational states in the ground state. 
    The last column gives the differences (Diff. = Obs. - Calc.) between the measured transition frequencies and those calculated using \textsc{pgopher} with the constants given in Table~\ref{tab:molecularConstants}. All frequencies are in MHz.}
    \label{tab41}
    \begin{tabular}{lllllllr}
    \hline
    \multicolumn{2}{l}{Lower state} & & \multicolumn{2}{l}{Upper state} & ~~Frequency & ~~Unc. & ~~Diff.\\
    \hline
\XS, $	N=0, J=1/2,	$ & $ 	F=1	$ &~~~ $\rightarrow$ ~~~& \Ab, $	J=3/2\,(-),	$ & $ 	F=1	$ ~~& ~~$ 	367632135.1		 $~~&~~$ 	0.3	 $~~&~$ 	-0.3	$ \\
$		$ & $ 	F=1	$ &~~~ $\rightarrow$ ~~~& $		$ & $ 	F=2	$ ~~& ~~$ 	367632154.2		 $~~&~~$ 	0.1	 $~~&~$ 	-0.2	$ \\
$		$ & $ 	F=0	$ &~~~ $\rightarrow$ ~~~& $		$ & $ 	F=1	$ ~~& ~~$ 	367632200.9		 $~~&~~$ 	0.2	 $~~&~$ 	-0.7	$ \\
\XS, $	N=1, J=3/2,	$ & $ 	F=2	$ &~~~ $\rightarrow$ ~~~& \Ab, $	J=5/2\,(+),	$ & $ 	F=2	$ ~~& ~~$ 	367650903.8		 $~~&~~$ 	0.5	 $~~&~$ 	-0.1	$ \\
$		$ & $ 	F=2	$ &~~~ $\rightarrow$ ~~~& $		$ & $ 	F=3	$ ~~& ~~$ 	367650916.4		 $~~&~~$ 	0.2	 $~~&~$ 	0.3	$ \\
$		$ & $ 	F=1	$ &~~~ $\rightarrow$ ~~~& $		$ & $ 	F=2	$ ~~& ~~$ 	367650938.0		 $~~&~~$ 	0.1	 $~~&~$ 	0.0	$ \\
\XS, $	N=2, J=5/2,	$ & $ 	F=3	$ &~~~ $\rightarrow$ ~~~& \Ab, $	J=7/2\,(-),	$ & $ 	F=4	$ ~~& ~~$ 	367669431.8		 $~~&~~$ 	0.7	 $~~&~$ 	0.4	$ \\
$		$ & $ 	F=2	$ &~~~ $\rightarrow$ ~~~& $		$ & $ 	F=3	$ ~~& ~~$ 	367669455.9		 $~~&~~$ 	0.3	 $~~&~$ 	0.4	$ \\
\XS, $	N=3, J=7/2,	$ & $ 	F=4	$ &~~~ $\rightarrow$ ~~~& \Ab, $	J=9/2\,(+),	$ & $ 	F=5	$ ~~& ~~$ 	367687700.4		 $~~&~~$ 	0.6	 $~~&~$ 	0.8	$ \\
$		$ & $ 	F=3	$ &~~~ $\rightarrow$ ~~~& $		$ & $ 	F=4	$ ~~& ~~$ 	367687726.0		 $~~&~~$ 	0.2	 $~~&~$ 	0.8	$ \\
\XS, $	N=4, J=9/2,	$ & $ 	F=5	$ &~~~ $\rightarrow$ ~~~& \Ab, $	J=11/2\,(-),	$ & $ 	F=6	$ ~~& ~~$ 	367705721.2		 $~~&~~$ 	0.7	 $~~&~$ 	0.6	$ \\
$		$ & $ 	F=4	$ &~~~ $\rightarrow$ ~~~& $		$ & $ 	F=5	$ ~~& ~~$ 	367705747.4		 $~~&~~$ 	0.5	 $~~&~$ 	0.3 $ \\
\XS, $	N=5, J=11/2,	$ & $ 	F=6	$ &~~~ $\rightarrow$ ~~~& \Ab, $	J=13/2\,(+),	$ & $ 	F=7	$ ~~& ~~$ 	367723493.7		 $~~&~~$ 	1.3	 $~~&~$ 	-0.5	$ \\
$		$ & $ 	F=5	$ &~~~ $\rightarrow$ ~~~& $		$ & $ 	F=6	$ ~~& ~~$ 	367723521.5		 $~~&~~$ 	2.0	 $~~&~$ 	0.1	$ \\
\XS, $	N=6, J=13/2,	$ & $ 	F=7	$ &~~~ $\rightarrow$ ~~~& \Ab, $	J=15/2\,(-),	$ & $ 	F=8	$ ~~& ~~$ 	367741019.0		 $~~&~~$ 	0.9	 $~~&~$ 	-1.0	$ \\
$		$ & $ 	F=6	$ &~~~ $\rightarrow$ ~~~& $		$ & $ 	F=7	$ ~~& ~~$ 	367741046.5		 $~~&~~$ 	1.3	 $~~&~$ 	-1.3	$ \\
    \hline
    \end{tabular}
\end{table*}

\begin{table*}
    \renewcommand{\tabcolsep}{1pt}
    \centering
    \caption{Measured transition frequencies between the \XS\ and the \AS\ manifolds starting from the lowest rotational level of different vibrational levels in the ground state. Differences (Diff. = Obs. - Calc.) relate to deviations from the \textsc{pgopher} computation with molecular constants listed in Tables~\ref{tab:molecularConstants} and \ref{tab:molecularConstantsv1}.
    All frequencies are in MHz. 
    }
    \label{tab42}
    \begin{tabular}{lllllllr}
    \hline
    \multicolumn{2}{l}{Lower state} & & \multicolumn{2}{l}{Upper state} & ~~Frequency & ~~Unc. & ~~Diff.\\
    \hline
\XS, $	v=0, N=0, J=1/2$ & $ 	F=1	$ &~~~ $\rightarrow$ ~~~& $	A^2\Pi_{1/2},v=0,J=1/2\,(-),	$ & $ 	F=0	$ ~~&~~$ 	348666402.6  $~~&~~$ 0.3$~~&$  0.3 $ \\
$		$ & $ 	F=1	$ &~~~ $\rightarrow$ ~~~& $		$ & $ 	F=1	$ ~~& ~~$ 	348666424.4 	 $~~&~~$ 	0.3	$~~&$  0.2 $ \\
$		$ & $ 	F=0	$ &~~~ $\rightarrow$ ~~~& $		$ & $ 	F=1	$ ~~& ~~$ 	348666490.0 	 $~~&~~$ 	0.3	$~~&$  -0.5 $ \\
\XS, $	v=1, N=0, J=1/2$ & $ 	F=1	$ &~~~ $\rightarrow$ ~~~& $	A^2\Pi_{1/2},v=1,J=1/2\,(-),	$ & $ 	F=0	$ ~~& ~~$ 	347728600.6		 $~~&~~$ 	0.2	$~~&$  -0.8 $ \\
$		$ & $ 	F=1	$ &~~~ $\rightarrow$ ~~~& $		$ & $ 	F=1	$ ~~& ~~$ 	347728622.3		 $~~&~~$ 	0.5	$~~&$  2.6 $ \\
$		$ & $ 	F=0	$ &~~~ $\rightarrow$ ~~~& $		$ & $ 	F=1	$ ~~& ~~$ 	347728687.5		 $~~&~~$ 	0.4	$~~&$  1.5 $ \\
\XS, $	v=1, N=0, J=1/2	$ & $ 	F=1	$ &~~~ $\rightarrow$ ~~~& $	A^2\Pi_{3/2},v=1,J=3/2\,(-),	$ & $ 	F=1	$ ~~& ~~$ 	366678862.6		 $~~&~~$ 	0.3	$~~&$  1.0 $ \\
$		$ & $ 	F=1	$ &~~~ $\rightarrow$ ~~~& $		$ & $ 	F=2	$ ~~& ~~$ 	366678881.9		 $~~&~~$ 	0.2	$~~&$  -0.2 $ \\
$		$ & $ 	F=0	$ &~~~ $\rightarrow$ ~~~& $		$ & $ 	F=1	$ ~~& ~~$ 	366678927.6		 $~~&~~$ 	0.2	$~~&$  -0.3 $ \\
\XS, $	v=2, N=0, J=1/2$ & $ 	F=1	$ &~~~ $\rightarrow$ ~~~& $	A^2\Pi_{3/2},v=2,J=3/2\,(-),	$ & $ 	F=1	$ ~~& ~~$ 	365722742.4		 $~~&~~$ 	0.4	$~~& \\
$		$ & $ 	F=1	$ &~~~ $\rightarrow$ ~~~& $		$ & $ 	F=2	$ ~~& ~~$ 	365722761.2		 $~~&~~$ 	0.2	$~~& \\
$		$ & $ 	F=0	$ &~~~ $\rightarrow$ ~~~& $		$ & $ 	F=1	$ ~~& ~~$ 	365722806.8		 $~~&~~$ 	0.7	$~~& \\
    \hline
   \end{tabular}
\end{table*}

\section{Analysis and Molecular constants}

For the analysis of the molecular structure an effective Hamiltonian for the \XS\ ground state is used~\cite{Ernst1986,Steimle2011}:
\begin{equation}
    H= B \vec{N}^2 + \gamma_{sr} \vec{N}\cdot \vec{S} + b\vec{I} \cdot \vec{S} + cI_zS_z
\end{equation}
where additional terms for centrifugal distortion to rotation and spin-rotation must be added, and $b$ and $c$ are the common Frosch and Foley parameters for the hyperfine interaction~\cite{Frosch1952}. The $b_F$ hyperfine constant~\cite{Steimle2011} equals $b_F=b-c/3$.
Only the $^{19}$F nucleus has a nuclear spin $I_F=1/2$, while the spin of the $^{138}$Ba nucleus is $I_{Ba}=0$.

For the analysis of the \AS\ excited state we follow the effective Hamiltonian formalism of Steimle et al.~\cite{Steimle2011}:
\begin{equation}
      H_{\rm eff} = T_{v',v''} + H_{\rm LS}  + H_{\rm rot} + H_{\Lambda} + H_{\rm hyp} 
\end{equation}
with
\begin{eqnarray}
\begin{split}
     H_{\rm LS}& = AL_zS_z + \tfrac{1}{2} A_D \{ L_zS_z, R \} \\
     H_{\rm rot} &=   B \vec{R}^2 - D[\vec{R}^2]^2 \\
     H_{\Lambda}  &= \tfrac{1}{2} (p+2q) (e^{-2i\phi}J_+S_+ + e^{+2i\phi}J_-S_- )
\end{split}
\end{eqnarray}
where the \{\} brackets indicate an anticommutator. 
This is congruent with the terms in \textsc{pgopher}.
The $\phi$ insert is to be treated as a phase factor referring to the signs in the equations to be considered, which then yields in a pure Hunds case (a) a $\Lambda$-doubling of $qJ^2$ in the \Ab\ ladder and $(p+2q)J$ in the \Aa\ ladder.
In the ensuing fitting procedures the state mixing between ladders is taken into account. In the actual fits, within the framework of  \textsc{pgopher}, a centrifugal distortion term in the $p_D$ $\Lambda$-doubling is also included.

Following the formalism of Denis et al.~\cite{Denis2022} for the hyperfine structure in the \AS\ excited state we adopt terms for both $\Lambda$-doublet ladders:
\begin{eqnarray}
\begin{split}
     H_{\rm hyp}(^2\Pi_{1/2}) &= [a-\tfrac{1}{2}(b+c)]I_z \\ 
     &+ \tfrac{1}{2}d (e^{+2i\phi}I^-S^- + e^{-2i\phi}I^+S^+) \\
     H_{\rm hyp}(^2\Pi_{3/2}) &=   [a+\tfrac{1}{2}(b+c)]I_z 
\end{split}
\end{eqnarray}
with $a$, $b$, $c$ and $d$ the Frosch and Foley parameters~\cite{Frosch1952}. 
In Ref.~\cite{Denis2022} it was verified that the hyperfine Hamiltonian and parameters were consistent with the definitions of \textsc{pgopher}.

The task of producing a fit to the effective Hamiltonians for the \XS\ and \AS\ states described above and extracting a set of molecular constants representing the measured transition frequencies to the level of their experimental accuracies is a complex one. 
The set of electronic transitions measured currently in the $A - X$ system is the most accurate in terms of absolute accuracy, but the number of targeted spectral lines is limited.
These pieces of information gained must be combined with the large amount of data from previous spectroscopic investigations. 

The level structure of the \XS($v=0$) ground state is best described by the results of previous experiments and the present data, while highly accurate, do not improve the molecular constants describing this state.
The hyperfine structure molecular constants $b$ and $c$ were investigated by using a fit to combine measurements using direct radio frequency excitations of many different $\Delta N$ transitions within $v=0$, delivering a highly accurate description~\cite{Ernst1986}. 
The rotational constants including high-order centrifugal constants $D$ and $H$, and the fine structure determined by the spin-rotation interaction constants $\gamma$, $\gamma_D$ and $\gamma_H$ are derived from the highly accurate microwave experiments~\cite{Ryzlewicz1980} and electronic transitions up to very high rotational quantum numbers~\cite{Effantin1990,Bernard1990}.
Even though the resolution in the Fourier-transform emission studies in the visible range were at a resolution of 0.027 \wn~\cite{Effantin1990}, the fact that rotational progressions were followed up to rotational quantum numbers exceeding 100 allowed for determining centrifugal distortion parameters to very high accuracy.

\begin{table*}
    \centering
    \caption{Molecular constants used in \textsc{pgopher}, a program for simulating rotational, vibrational and electronic spectra~\cite{Western2017}. Constants are taken from Refs.~\cite{Ryzlewicz1980,Ernst1986,Effantin1990,Bernard1990,Steimle2011,Denis2022}, except for the state origin of the \AS$(v=0)$ manifolds and the spin-orbit coupling constant $A$ in the \AS$(v=0)$ state, which are fitted (with all other constants fixed) using \textsc{pgopher} on the transitions listed in Table~\ref{tab42}. 
    }
    \label{tab:molecularConstants}
    \begin{tabular}{lllll}
    \hline
State &    Molecular constant & Symbol & Value (cm$^{-1}$) & Value (MHz)\\
\hline  
\XS$(v=0)$ &  Vibrational term value   &  $T_0(X) $  & $0.0$  &  $0.0$ \\
& Rotational constant & $B$ & $ 0.21594802 $ & $ 6473.9588^a $ \\
& Quartic centrifugal distortion & $D$ & $ 1.84294\times10^{-7~b}$ & $ 0.0055250 $ \\
& Sextic centrifugal distortion & $H$ & $ 1.40\times10^{-14~b} $ & $ 4.2\times10^{-10} $ \\
& Spin rotation coupling constant & $\gamma$ & $ 0.0027013 $ & $ 80.984^a $ \\
& Centrifugal distortion of $\gamma$ & $\gamma_D$ & $ -1.95\times10^{-6} $ & $ -0.0584^a $ \\
& $J^6$ Centrifugal distortion of $\gamma$ & $\gamma_H$ & $ 3.7\times10^{-9} $ & $ 0.000112^a $ \\
& Magnetic hyperfine parameter & $b$ & $ 0.0021184 $ & $ 63.509^e$ \\
& Magnetic hyperfine parameter & $c$ & $ 0.0002743 $ & $ 8.224^e $ \\
\AS$(v=0)$ & Band origin & $ T_{0,0}$ & $ 11946.109609\,(7) $ & $ 358135356.32\,(21)^c $ \\
& Rotational constant & $B$ & $ 0.2117414^d $ & $ 6347.847 $ \\
& Quartic centrifugal distortion & $D$ & $2.0036 \times10^{-7~b}$ & $0.006007$\\
& Spin-orbit coupling constant & $A$ & $ 632.287838\,(13) $ & $ 18955512.5\,(4)^c $  \\
& Centrifugal distortion of $A$ & $A_D$ & $ 3.1\times10^{-5~d} $ & $ 0.93 $ \\
& $\Lambda$-doubling constant & $p$ & $ -0.257310 $ & $ -7713.96^b $ \\
& $\Lambda$-doubling constant & $q$ & $ -8.40\times10^{-5} $ & $ -2.52^b $ \\
& Centrifugal distortion of $p$ & $p_D$ & $ -2.332\times10^{-7} $ & $ -0.00699^b $ \\
& Magnetic hyperfine parameter & $a$ & $ 0.0008856 $ & $ 26.55^f $ \\
& Magnetic hyperfine parameter & $b+c$ & $ -0.000185 $ & $ -5.54^f $ \\
& Magnetic hyperfine parameter & $d$ & $ 0.000119 $ & $ 3.58^f $ \\
    \hline
    \multicolumn{4}{l}{\footnotesize{$^a$Ryzlewicz \emph{et al.}~\cite{Ryzlewicz1980}}}\\
    \multicolumn{4}{l}{\footnotesize{$^b$Effantin \emph{et al.}~\cite{Effantin1990}}}\\
    \multicolumn{4}{l}{\footnotesize{$^c$This work}}\\
    \multicolumn{4}{l}{\footnotesize{$^d$Steimle \emph{et al.}~\cite{Steimle2011}}}\\
     \multicolumn{4}{l}{\footnotesize{$^e$Ernst \emph{et al.}~\cite{Ernst1986}}}\\
    \multicolumn{4}{l}{\footnotesize{$^f$Denis \emph{et al.}~\cite{Denis2022}}}
    \end{tabular}
\end{table*}

\begin{table*}
    \centering
    \caption{Molecular constants for the \XS$(v=1)$ and \AS$(v=1)$ vibrational states.
    }
    \label{tab:molecularConstantsv1}
    \begin{tabular}{lllll}
    \hline
State &    Molecular constant & Symbol & Value (cm$^{-1}$) & Value (MHz)\\
    \hline
\XS$(v=1)$ & Vibrational term value & $T_1(X)$  & $ 465.745446\,(66)^a$  &  $13962697.2\,(2.0)$ \\
& Rotational constant & $B$ & $ 0.214785499^a $ & $6439.10727$ \\
& Quartic centrifugal distortion & $D$ & $ 1.8496\times10^{-7, a}$ & $0.0055390$ \\
& Spin rotation coupling constant & $\gamma$ & $ 0.002713^b $ & $81.33$ \\
& Magnetic hyperfine parameter & $b$ & $ 0.0021184 $ & $ 63.509^c $ \\
& Magnetic hyperfine parameter & $c$ & $ 0.0002743 $ & $ 8.224^c $ \\
\AS$(v=1)$ & Band origin & $T_{1,1}$  & $12380.3184805\,(8)$ & $371152610.81\,(25)$  $^d$ \\ 
& Vibrational term value & $ T_1(A)$ & $12846.063926\,(66)  $ & $385115308.0\,(2.0)$  $^d$   \\
& Rotational constant & $B$ & $ 0.2105271^b $ &  $6311.444$\\
& Quartic centrifugal distortion & $D$ & $ 2.0059\times10^{-7~b} $ & $0.0060135$ \\
& Spin-orbit coupling constant & $A$ & $ 631.722697\,(17) $ & $18940069.0\,(5)^d$\\
& Centrifugal distortion of $A$ & $A_D$ & $ 5.025\times10^{-5~b} $ &  $1.5064$ \\
& $\Lambda$-doubling constant & $p$ & $ -0.257941^b $ & $-7732.88$ \\
& $\Lambda$-doubling constant & $q$ & $ -7.6\times10^{-5~b} $ & $-2.28$ \\
& Centrifugal distortion of $p$ & $p_D$ & $ -2.25\times10^{-7~b} $ &  $-0.00674$\\
& Magnetic hyperfine parameter & $a$ & $ 0.0008856 $ & $ 26.55^e $ \\
& Magnetic hyperfine parameter & $b+c$ & $ 0.000185 $ & $ -5.54^e $ \\
& Magnetic hyperfine parameter & $d$ & $ 0.000119 $ & $ 3.58^e $ \\
    \hline
    \multicolumn{4}{l}{\footnotesize{$^a$Guo \emph{et al.}~\cite{Guo1995}}}\\
    \multicolumn{4}{l}{\footnotesize{$^b$Effantin \emph{et al.}~\cite{Effantin1990}}}\\
   \multicolumn{4}{l}{\footnotesize{$^c$Kept fixed to the values for \XS$(v=0)$ from Ernst \emph{et al.}~\cite{Ernst1986}}}\\ \multicolumn{4}{l}{\footnotesize{$^d$This work}}\\
    \multicolumn{4}{l}{\footnotesize{$^e$Kept fixed to values for \AS$(v=0)$ from Denis \emph{et al.}~\cite{Denis2022}}}
    \end{tabular}
\end{table*}

The same is true for the \AS($v=0$) excited state. The rotational constants and their centrifugal distortions, and the $\Lambda$-doubling parameters were determined from the sequences of rotational branches in the $A-X$ system up to high $J$. 
The fact that the molecular constants describing these rotational and $\Lambda$-doubling effects scale with $J^2$ or even higher order in $J$ makes that the present data do not deliver improvement.
The analysis of the hyperfine structure in the $A-X$ system can also be separated from the overall rovibronic structure of the excited state. 
The hyperfine structure in BaF was analyzed previously in a combined theoretical and experimental study and values for the Frosch and Foley parameters for the excited \AS\ state of $^{138}$Ba$^{19}$F were derived~\cite{Denis2022,Marshall2025}. 

Based on this logic, a fit to the present data is performed using the \textsc{pgopher} program~\cite{Western2017}. 
The resulting list of molecular constants describing the $A-X$(0,0) band is presented in Table~\ref{tab:molecularConstants}.
By keeping the molecular constants describing rotation, spin-rotation interaction in the ground state, as well as $\Lambda$-doubling  and the rotational constant and centrifugal distortions in the excited state fixed in a comprehensive fit, it is ascertained that the rotational structure remains accurately described by the previously determined set of constants. 
We note that it is of particular importance to keep the rotational constant and centrifugal distortions fixed to the work of Ref.~\cite{Effantin1990}because that was derived from rotational levels up to and beyond $J > 100$.
Also the hyperfine structure in ground and excited states was kept at the accurate analyses of Refs.~\cite{Ernst1986,Denis2022}.

Effectively, in the fit only the band origin $T_{0,0}$ and the spin-orbit constant $A$ are determined. 
The high absolute accuracy of the transition frequencies, derived from the narrow lines calibrated by the frequency comb standard, is transferred into these two molecular constants while keeping the rotational structure as is, thus not affecting the previous spectroscopic analyses~\cite{Ryzlewicz1980,Effantin1990}.
Deviations are found between currently determined values and those of Ref.~\cite{Steimle2011}. The difference for the $T_{0,0}$ band origin amounts to 0.21 \wn, corresponding to the rotational constant. So this may be attributed by different definitions in the effective Hamiltonian. This may also hold for the deviations for the $A$ spin-orbit constant, which corresponds in fact to the difference between band origin values. The small difference of 0.006 \wn\ may correspond to a differential effect of rotational energies included in the band origins.

The resulting fitted values are found to be in good agreement with the experimental frequencies measured in the $A-X$(0,0) band, with deviations (listed in Tables~\ref{tab41} and \ref{tab42}) on average smaller than 1 MHz.
The good agreement between experiment and results from the fit for the higher-$J$ values in the $^SR$-branch, as in Table~\ref{tab41}, demonstrate that the neglect of the contribution by the weaker $\Delta J=1, \Delta F=0$ hyperfine components is justified.

The dataset probing transitions in the $A-X$(1,1) band is limited to the excitation of the lowest rotational levels in both the \Aa\ and \Ab\ spin-orbit ladders.
This set of two independent measurements (with hyperfine substructure) allows, in the same manner as for the (0,0) band, the determination of the two molecular constants setting the absolute accuracy, the band origin $T_{1,1}$ and the spin-orbit constant $A$. These were determined in a separate fit using \textsc{pgopher}, while keeping all other molecular constants fixed, similar and for the same reasons as for the (0,0) band.
Neither in the \XS($v=1$) ground level nor in the \AS($v=1)$ excited state a set of hyperfine constants has been determined. For this reason we keep the hyperfine constants for ($v=1$) levels the same and fixed at the values for ($v=0$).
Inspection of the measured splittings in the two lines for the (1,1) band as listed in Table~\ref{tab42} shows that indeed the hyperfine splittings in \XS($v=1$) are the same (within error margins)  to those in \XS($v=0$).
The same holds for the hyperfine splittings in the \AS($v=1$) vibrational level, being similar to those in \AS($v=0$).

Results from the \textsc{pgopher} fit for the $A-X$(1,1) band are presented in Table~\ref{tab:molecularConstantsv1}.
In order to determine the excitation energy of the \AS($v=1$) level the vibrational excitation energy $T_1(X)$ in the ground state, obtained from the infrared FT-emission study~\cite{Guo1995}, is added to determine the excitation energy $T_1(A)$ of the first vibration in the excited state.
The deviations between observed and calculated transition frequencies based on the \textsc{pgopher} fit are at the MHz level, as displayed in Table~\ref{tab42}.
The excitation energy of the \AS($v=1$) vibrational level is obtained by adding the \XS($v=1$) term value, as obtained from the FT-infrared emission studies~\cite{Guo1995}, to the presently determined band origin $T_{1,1}$. The resulting uncertainty of 2.0 MHz is determined by the limited accuracy of the FT-infrared emission work.

\section{Conclusion}

A high-resolution spectroscopic study is performed on the $A-X$ BaF system, probing both spin-orbit levels \Aa\ and \Ab, using a cold molecular beam set-up involving a buffer gas cell as the beam source.
Novel to the study is the full resolution of hyperfine structure and the absolute calibration with a frequency comb laser delivering sub-MHz accuracy.
A set of improved molecular constants for the \AS\ state is derived, at higher accuracy than in previous studies. 
Besides a study on the (0,0) band molecular constants for the (1,1) band are also derived.
In particular, highly accurate values for the band origins and spin-orbit constants in the \AS($v=0$) and ($v=1$) levels are obtained, both of relevance for laser-cooling schemes of BaF molecules.
The extracted molecular constants reproduce the measured transition frequencies at the experimental absolute accuracy of 1 MHz.
The quantitative spectroscopic information is of relevance for the application of laser-cooling of BaF molecules, which is currently being explored in the NL-$e$EDM consortium~\cite{vanHofslot2025}, and in a broader sense important for future $e$EDM experiments based on the BaF target system.

\section*{Data Availability}
The data on transition frequencies are contained in the \textsc{pgopher} file made available as Supplementary Material. Further data underlying the spectra can be made available upon reasonable request.

\section*{Supplementary material}
The \textsc{pgopher} files containing all transition frequencies and the fits determining the molecular constants represented in Tables~\ref{tab:molecularConstants} and \ref{tab:molecularConstantsv1} are made publicly available in digital form.

\section*{Acknowledgements}
The NL-eEDM consortium receives program funding (EEDM-166 and XL21.074) from the Netherlands
Organisation for Scientific Research (NWO). 
The authors thank an anonymous referee for indicating an explanation for the observed intensity anomalies.

\bibliography{BaF-bib}

\end{document}